

\parindent 0pt
\magnification=1200
\null

\vskip 0.60in
\centerline{Glueball Spectra of SU(2) Gauge Theories in 3 and 4 Dimensions:}
\vskip 0.15in
\centerline{A Comparison with the Isgur-Paton Flux Tube Model}
\vskip 0.90in
\centerline{T. Moretto}
\vskip 0.2in
\centerline{NORDITA} 
\centerline{Blegdamsvej  17}
\centerline{DK-2100 Copenhagen \O, Denmark}
\vskip 0.25in
\centerline{and}
\vskip 0.25in
\centerline{M. Teper}
\vskip 0.2in
\centerline{All Souls College and Department of Physics}
\centerline{University of Oxford} 
\centerline{1 Keble Road}
\centerline{Oxford OX1 3NP, UK}

\vskip 1.0in
$\rm{\underline{Abstract}} \ \ \ $ We use the results of recent
lattice calculations to obtain (part of) the
mass spectrum of continuum SU(2) gauge theory in
both 2+1 and 3+1 dimensions. We compare these spectra to the
predictions of the Isgur-Paton flux tube model for glueballs.
We use this comparison to test the reliability of different
aspects of the model and also to learn which aspects  of the
lattice calculations it is important to improve upon.

\vfill\eject
\parskip 10pt plus 1pt minus 1pt

Current lattice calculations$^{(1-4)}$ of the SU(2) spectrum in 3+1
dimensions are accurate, have a good control  of finite-volume
corrections,  and have been performed down to very small values
of the lattice spacing, $a$. This should encourage us to see 
whether an accurate extrapolation to the continuum limit is
possible. Similar calculations$^{(5)}$ in 2+1 dimensions are even
more accurate and some such extrapolations have already been
made in that case.

In both dimensions the theories are believed to possess  linear 
confinement at large distances and they become free at short distances.
These similarities suggest that a comparison of the respective
mass spectra should be informative. However there is a limit as to
how focussed our comparison of the spectra can be if posed in
the abstract. To do better it will clearly be very useful to have
a model framework within which to couch the discussion. We can then 
hope to pin-point  which dynamical features of the model work
in practice  and which do not. Using two different spatial
dimensions provides an extended  lever arm for the comparison.
In return we can expect that the model will focus our attention
on what are the weak points of current lattice calculations.

Glueball models are more speculative than models for the usual
hadrons, where at least the basic starting point, valence
(constituent) quarks, is well established. We are aware of two
models that might be useful for our purposes here: the bag
model$^{(6)}$ and the flux tube model$^{(7)}$. In this paper
we shall work with the flux tube model. A detailed comparison
with the bag model would also be of interest. However in that 
case there is a complication - the lightest $0^+$ state has 
an imaginary mass once the centre of mass motion
is subtracted - and it is not clear to us to what extent the
usual resolution of this problem should be considered uncontroversial.

We turn now to the lattice glueball spectra. Lattice masses
always come in lattice units and so, to remove the lattice
spacing, we will consider ratios of masses. In general the
most accurately calculated physical quantity is the string
tension, $\sigma$. Therefore the mass ratios that we shall 
extrapolate to the continuum limit will be of the form
$m/\surd\sigma$ where $m$ is a glueball mass. Now the
leading lattice correction to such a mass ratio is known
to be of the form$^{(8)}$ $O((a/\xi)^{2})$ where $\xi$ is some
physical length scale. We choose to use the length scale
$\xi_{\sigma}=1/\surd\sigma$ in which case this leading
correction is proportional to $a^{2}\sigma$. (This correction may
also contain a dependence on $g^2$ but this varies
weakly, if at all, with $a$ and so we neglect it.) We now fit the 
D=4 lattice results$^{(1-4)}$ in the range $\beta \equiv 4/g^{2} \le 2.85$ 
with a formula that incorporates this correction and we obtain the
continuum mass ratios shown in the relevant column of 
Table 1. For D=3 we perform an
identical analysis on the results of ref(5), for the range
$\beta \equiv 4/ag^{2} \le 14.5$, and we obtain the second 
column of Table 1.

To  be confident in these results requires a rather detailed
analysis of the mass calculations at each value of $\beta$.
This will be provided in a longer paper$^{(9)}$. That paper
will also contain a careful description of the flux-tube
model, which we only sketch here, and it will contain
a much more comprehensive comparison of spectra than we are
able to provide in this Letter.

One point of detail that we are forced to address here is the
spin assignment. The glueball wave-functionals are constructed
using only the lattice rotational symmetries. Thus a state
that we call $J=0$ may actually be $J=4,8,..$. The usual
practice is to label the state by the lowest allowed value
of $J$ on the assumption that that is presumably the lightest
of the contributing states. Our analysis of the flux-tube
model will show that this assumption  breaks down too often
for it to be really useful. Fortunately one can easily
show$^{(9)}$ that where the basic components used in constructing
the glueball wavefunctionals are broad and smooth - as are
the smeared operators used in those lattice calculations whose
results we employ - the lattice operator will have its largest
individual projection onto the state with the lowest allowed $J$.
One can therefore infer that when we extract
the lightest mass contributing to the correlation function
of such smeared operators, if the normalised projection of the 
operator onto this lightest mass state is much greater
than 0.5, then it is extremely unlikely that that state
possesses anything other than the lowest allowed value of $J$.
By this criterion all the spin assignments in Table 1
are correct and should be taken at face value.

Before turning to the model it is interesting to compare
the two spectra in Table 1. What is most striking is
how similar they are, taking into account that in 3 dimensions
$J\ne0$ states of opposite parity should be degenerate - 
which is what we observe (within errors). 
For both dimensions the lightest state is 
the $0^+$ and the next lightest is the $2^+$, with the
$2^-$ and $0^-$ not too far away. Then come the heavier
states of spin $J=1,3$. Even the ratios of the $2^+$ to
$0^+$ masses are not very different. The  $0^+$ glueball is
heavier in D=3 than D=4 when expressed in units of the string
tension and we might wonder whether this can be simply related
to the change in the number of spatial dimensions. As we shall
see below, the flux-tube model does provide an
explanation  of this kind.

In theories with linear confinement one expects the flux
between well separated fundamental sources to be localised in a 
stringlike flux-tube. If one considers a flux-tube  that
closes upon itself rather than ending on quarks, then we
have a colourless, quarkless `excitation' which we can 
imagine providing us with the basic component of a
glueball state. This is the starting point of the Isgur-Paton 
flux-tube model$^{(7)}$. 

We shall begin with the D=2+1 case. Consider as our starting point
a circular flux-loop of radius $\rho$ lying in the 
spatial plane. If we ignore the internal structure of the loop, then
its oscillations are of two types. First there are the periodic
oscillations about this circle. Secondly there are the `collective'
radial oscillations of the whole loop. The oscillations of the
first type, upon quantisation, are 
described by phonons, of frequency $m/\rho$. We can define them so
that they have angular momentum $\pm m$. Let $n^{+}_{m},n^{-}_m$ be the
number of phonons  with frequency $m/\rho$ and with $J=\pm m$
respectively. Then the
total angular momentum contributed by the phonons is

$$ J = \sum_{m} m\left(n^{+}_m - n^{-}_m\right) \eqno(1)$$

and the total excitation energy will  be $E=M/\rho$ where 

$$ M = \sum_{m} m\left(n^{+}_m + n^{-}_m\right) \eqno(2)$$

In D=3 parity flips angular momentum
so it corresponds to interchanging the $+$ and $-$ phonons.
Now, infinitesimal $m=1$ oscillations are easily seen to be equivalent
to infinitesimal translations and rotations. Thus we shall follow
ref(7) and exclude these modes from consideration. This is analogous to
the cm momentum subtraction in the bag model that we
referred to earlier. Since these are simple harmonic oscillators
they contribute a zero point energy, and this clearly diverges.
The divergent piece is proportional to the string length and
can be absorbed into the bare string tension to produce the
observed string tension, $\sigma$. For long strings the leading
remaining piece is simply $13\pi(D-2)/6l$ where $l$ is the string
length $2\pi\rho$. We recognise this
as being the usual L\"uscher universal string correction$^{(10,11)}$ 
with the contribution of the $m=1$ mode removed. Finally we put in
a factor to soften the $1/\rho$ behaviour because we know
that the string is really a flux-tube with a width of order
$1/\surd\sigma$. Putting all this together we can write the
total energy of the string as 

$$ E^{M}_{s}\left(\rho\right) = 
2\pi\rho\sigma + {{M-\gamma}\over{\rho}}
\left(1-e^{-f\surd\sigma\rho}\right) \eqno(3)$$

with M related to the phonon content of the string by eqn(2). While 
we shall keep in mind both that
the theoretically preferred value
of $\gamma$ is the string value, $13(D-2)/12$, and 
that we expect $f\sim 1$, we shall treat both
of these as free parameters in our calculations below
(again following ref(7)).

The second type of oscillation  consists of variations in
$\rho$. If we think of this as being a collective `slow'
oscillation, then we can simplify the problem of quantising
the flux-tube by making an adiabatic approximation$^{(7)}$ where
these `slow' oscillations take place in a potential
provided by the `fast' phonon oscillations. Then the
energy eigenstates of the flux loop are characterised
by the numbers of phonons of various types and a radial
wavefunction that is a solution of the radial Schrodinger
equation

$$ \{ {{-9}\over{16\pi\sigma}}{{d^2}\over{d\xi^2}}
+ E^{M}_{s}(\xi^{3/2}) \} \psi(\xi) =
E \psi(\xi) \eqno(4)$$  

using the variable $\xi = \rho^{3/2}$ which turns out$^{(7)}$ to 
be what is appropriate here. 
We solve this equation for the energies and corresponding 
eigenstates using the numerical method of ref(12). 

Before turning to the detailed glueball mass spectra thus obtained, it
is worth reconsidering the adiabatic approximation used above.
The model, like the theory, possesses only an overall mass scale
which we may choose to be $\surd\sigma$. In units of this  mass
scale the whole spectrum is then fixed (for fixed values of $\gamma$
and $f$). So there is no obvious `limit' where the adiabatic
assumption becomes manifestly accurate. This is unlike the case with
quarks where we can consider the limit of large quark masses. 
It is more-or-less obvious that in such a one-scale problem we
should not expect the adiabatic assumption to be very good and
so it would be pointless to look for a perfect fit between
the model predictions and the real world of Table 1. What we
should rather do is to identify those parts of the spectrum where 
there is reasonable agreement and to try and relate
any (dis)agreement to particular aspects of the model's dynamics.

We can also test the adiabatic assumption self-consistently.
If it held very well then we would normally expect the splittings 
due to the phonon excitations to be much larger than 
the splittings due to
radial excitations. In Fig.1 we show a typical example of 
the masses of states of various radial, $n$, and phonon, $M$,
quantum numbers.  We see that the phonon and radial excitations
lead to similar energy splittings. This tells us that the
adiabatic approximation, while not necessarily very poor, is
likely to be rather crude. A similar situation is found to hold
in 4 dimensions. In fact in that case we have$^{(13)}$, from ref(4),
a mass estimate for the first $0^+$ excited state and it is
very close in mass to the lightest $2^+$, just as predicted
by the model. (This is a result at one value of $\beta$ and
so has to be taken cautiously.)

We show in Fig.2 the lowest states of the model as functions
of the parameter $f$ and for two values of $\gamma$
(including the theoretically favoured value). The lightest
state is the circular loop with no phonons which is clearly
$0^+$. The phonon states  provide the $J \ne 0$ states which
are parity doubled. On each of these states we have a tower
of radial excitations with the same spins and parities. Since
we exclude phonons with $m=1$ it is clear that the first
$J \ne 0$ states above the $0^+$ will have $J=2$, from one
$m=2$ phonon. To obtain $J=1$ we need a phonon content of, for
example, $n^{+}_{3}=1$ and $n^{-}_{2}=1$; and so the state will
be considerably heavier. All this is qualitatively like the actual
spectrum summarised in Table 1. Moreover, as we see in Fig.2,
the predicted value of $m(0^{+})/\surd\sigma$ is quite close to
its observed value, if we choose $f$ near its expected value of 
unity. However the $0^-$  glueball, which requires large $M$,  
e.g. $n^{+}_{4}=1$ and $n^{-}_{2}=2$, is predicted to
be considerably heavier than the $J=1$ states and this is
contrary to Table 1. Indeed it is only near $f=1$ that the
$0^{+},0^{-}$ states are in quantitative agreement with
the spectrum in Table 1 and in that case the  $J=1,2$
states are much too light. However if we look$^{(9)}$ at the
wave-functions of the states we see that the $J=0$ states
in particular are localised near small $\rho$ - which is why
they are so sensitive to the value of $f$ at small $f$ -
and for such states the string picture is likely to be
least reliable. Which brings us back to our earlier 
observation that it surely makes no sense to be searching
for an exact agreement between the model and the real world.

In Fig.2 we also show the predictions of the model for
the lightest $J=3$ and $J=4$ states. We observe that 
the $4^-$ is lighter than the $0^-$ and that the $J=3$ 
state is lighter than the $J=1$ state. This
highlights how dangerous is the usual assumption that
underlies the choice of labelling of states calculated 
on the lattice. The fact that the predicted $4^-$ is
so close to the observed `$0^-$', makes the question of 
whether, in fact, this latter state might not be $4^-$ 
particularly important. As we said earlier, we  believe
that there are convincing arguments to show that 
this is not the case.

We now turn to the model in 4 dimensions. There are two
additional types of oscillation. The first consists of
periodic vibrations orthogonal to the plane of the loop.
This leads, upon quantisation, to an additional kind of
phonon. The zero-point energy is doubled and so the
theoretically favoured value of $\gamma$ becomes
$13/6$ rather than $13/12$ as in 3 dimensions.
The second additional type of oscillation is associated
with the collective rotation of the plane of the loop. In the model
this is assumed to be a `slow' fluctuation and hence
one that enters the Schrodinger equation through the
addition of a familiar angular momentum term. The
states produced purely from phonons will still possess
parity doubling and it is these collective angular
excitations that allow the D=4 spectrum not to be
parity doubled for many values of $J$. If the adiabatic
approximation was  very good then the lightest $J\ne 0$
states would consist of angular excitations
of the $0^+$ ground state and we would expect much
smaller energy splittings than  in D=3. However in
practice the angular excitation energy is comparable to the
basic phonon excitation energy, just as we found for radial 
excitations, and so it is not clear how these new
excitations will alter the spectrum. 

We show in Fig.3 the  calculated spectrum in 4 dimensions
as a function of $f$ and for two values of $\gamma$, including
the theoretically favoured value of $\gamma = 13/6$. 
For values of $f$ close to unity the spectrum resembles, at
least qualitatively, the observed spectrum. That is, the lightest
state is a $0^+$ with the $2^{+},2^{-},0^-$ states next, and
$J=1,3$ forming a heavier cluster of states. Indeed at $f\simeq 1$
the agreement is quite good even at the quantitative level. In
D=4 the two kinds of phonons allow us to have a lighter $0^-$
than in D=3. A striking prediction of the model that seems to
hold for all values of $f$ and $\gamma$ is that the $J=3$
states are parity-doubled, unlike the $J=1$ states that are 
somewhat heavier in the model. Unfortunately these 
states possess large errors in the current lattice
calculations and so we cannot test these predictions
at present.

We return now to our earlier observation that the mass ratio
$m(0^{+})/\surd\sigma$ is larger by about $25\%$ in D=3 than
in D=4. As we see in Table 1 there seems to be some such upward
rescaling for the low-lying spectrum as a whole. It turns out
that this phenomenon has a very simple origin in the flux-tube 
model. This can be seen if we compare the D=3 and D=4 spectra
for the string values of $\gamma$, as in Fig.2b and Fig.3b
respectively. We observe that the $0^+$ mass ratio is 
indeed larger in D=3 and that for $f\simeq 1$ the values
are close to those in Table 1. If on the other hand we
compare the D=3 and D=4 spectra at the $same$ value of
$\gamma$ we see very little difference. Thus the origin
of the difference lies in the factor of two difference
between the values one calculates for $\gamma$ in 3 and
4 dimensions. This arises from a factor of two
between the zero-point energies in 3 and 4 dimensions. This
in turn follows from the fact that in 3 spatial dimensions
there are twice as many types of transverse  fluctuation
for the string as compared to 2 spatial dimensions.
So the point is that the zero-point energy reduces the energy 
density of a string of finite length, as in a glueball,
from the value of $\sigma$, and the reduction is twice as great 
in D=4 because there are twice as many transverse directions
in which the string can oscillate. This argument is a rigorous
one for the lowest-lying $0^+$ state because for that 
state it is only through the value of $\gamma$ that the 
dimensionality enters the calculation. Moreover it is clear
that this is an effect that will apply
to any glueball state that is composed of a small flux loop.

The flux-tube model does surprisingly well in reproducing the overall
features of the spectra, in both 3 and 4 dimensions. Perhaps its main
weakness is that it predicts the $0^-$  to be always heavier than
the $1^{\pm}$ in D=3.  On the other hand one might regard the
fact that the $J=1$ states are heavier than the $J=2$ states
as a surprise and so the fact that the model reproduces this
feature is a success for it. In the model it
arises from the absence of phonons with $m=1$ and from the
fact that the flux loop has no direction for SU(2). The latter
is no longer the case for SU(3) and will pose a problem for the 
model because the D=4 lattice SU(3) spectra do not show any light
$J=1$ states. (However we leave the SU(3) comparison till
the D=3 spectra become available.) 

In the model it is the phonon excitations that naturally produce 
parity doubling. So the D=4 spectrum will display parity doubling 
for those quantum numbers where the phonon states are the lightest
ones - such as $J=3$. This prediction tests the fundamental
dynamical assumptions of the model. It is unfortunate that
the lattice results are still too poor, for such heavier
glueballs, to test this prediction. Of course the basic
assumption of the model is that glueballs are essentially
finite loops of flux. One may view it as significant that
this provides a very simple explanation for the observed
fact that, when expressed in units of the string tension,
the lowest lying glueballs are lighter in 4 than in 3 dimensions. 

The comparisons with the flux-tube model have focussed
attention on features of the spectrum that might otherwise 
have appeared to be without any special significance. In 
particular it has focussed attention on some things
that lattice calculations need to do better. For example
it is crucial that we be confident in the
continuum spin assignments for states  on the lattice; and
it is also important to calculate the masses of states such
as the $J=4$ etc. Indeed the time has come to construct
lattice wave-functionals that are good approximations, up
to corrections of $O(a^{2})$  say, to states of a particular
continuum $J$. There is no real difficulty in doing so.
It is also clear that it is not only the lowest-lying states
of any given $J^P$ that are important; the excitations
carry valuable information, in this model, about the
splittings due to the `collective' radial and angular excitations. 
The energies of such states are easily calculable in lattice
calculations although it has usually not been thought important
to do so and almost no published results exist. Finally
heavier states, such as the  $J=3$ states in 4 dimensions,
can also  be important, as was pointed out in the previous
paragraph. On the lattice these are most easily calculated
by working with small lattice spacings. This will be 
time-consuming but not outrageously so.

\vskip 0.50in
$\rm {\bf{Acknowledgements}}$

We are grateful to Jack Paton for very helpful discussions.
We acknowledge the Oxford Particle Theory Grant SERC GR/F/41501.
One of us (TM) is grateful to the Danish Natural Science Research
Council for financial support during part of this work.

\vskip 0.70in

$\bf{\underline{References}}$

1) C.Michael,M.Teper: Phys.Lett.199B(1987)95;Nucl.Phys.B305(1988)453.

2) C.Michael,G.Tickle,M.Teper: Phys.Lett.207B(1988)313.

3) C.Michael,S.Perantonis: J.Phys.G18(1992)1725.

4) UKQCD Collaboration: Nucl.Phys.B394(1993)509.

5) M.Teper: Phys.Lett.289B(1992)115;311B(1993)223; and in progress.

6) T.H.Hansson: in  $Non-Perturbative Methods$ (Ed. S.Narison,
World Scientific, 1985)

7) N.Isgur,J.Paton: Phys.Rev.D31(1985)2910.

8) K.Symanzik: Nucl.Phys.B226(1983)187.

9) T.Moretto: Oxford D.Phil.Thesis,1993; T.Moretto,M.Teper: in progress.

10) M.L\"uscher,K.Symanzik,P.Weisz: Nucl.Phys.B173(1980)365.

11) Ph.de Forcrand,G.Schierholz,H.Schneider,M.Teper: Phys.Lett.160B(1985)137.

12) K.Miller,M.Olsson: Phys.Rev.D25(1982)2383. 

13) C.Michael: private communication.

\vskip 0.40in

$\bf{\underline{Figure \ Captions}}$

$\rm{\underline{Fig.1}}$: The lowest few $0^{+,-}$ and $2^{+,-}$
states as functions of $f$ in 3 dimensions. $M$ is the
phonon quantum number and $n$ is the radial quantum number. The points
display the actual calculated spectra and the lines are just to
guide the eye. Dashed lines connect radially excited states.

$\rm{\underline{Fig.2}}$: The D=3 glueball spectrum as a function
of $f$ and for the values of $\gamma$ indicated. The column of  
points at the far right comes from Table 1. Other points display
the values obtained with the model. The lines are to guide the
eye. The dashed line is for the $0^-$.

$\rm{\underline{Fig.3}}$: The D=4 glueball spectrum as a function
of $f$ and for the values of $\gamma$ indicated. The column of  
points at the far right comes from Table 1. Other points display
the values obtained with the model. The lines are to guide the
eye. Full lines connect positive  parity states, dashed lines
negative parity.

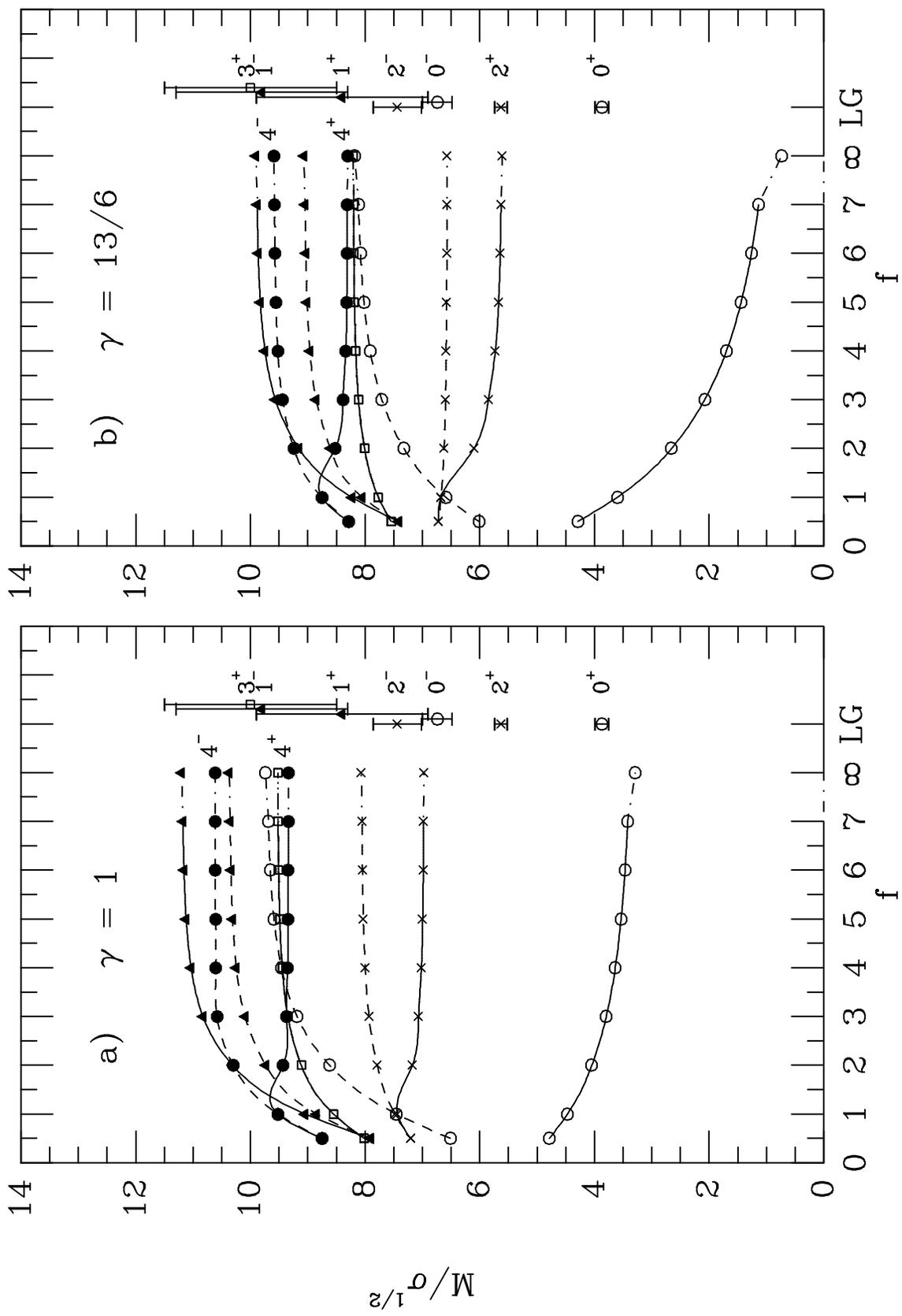
\end